\documentstyle[11pt,paspconf,epsf]{article}

\begin{document}

\title{Time Delay of PKS1830--211 Using Molecular Absorption Lines}

\author{Tommy Wiklind}
\affil{Onsala Space Observatory, SE-43992 Onsala, Sweden}

\author{Fran\c{c}oise Combes}
\affil{Observatoire de Paris, 61 Av. de l'Observatoire, F--75014 Paris,
France}

\begin{abstract}
The use of molecular absorption lines in deriving the time delay in
PKS1830--211 is described, as well as results from a three year
monitoring campaign. The time delay and the implied
value for the Hubble constant are presented.
\end{abstract}


%\section{Monitoring of PKS1830--211}
The gravitationally lensed radio source PKS1830--211 contains
two components; the NE and SW components, each consisting of
a core and a jet--like feature (cf. Subrahmanyan et al. 1990; Jauncey
et al. 1991). The separation of the core components is 0.98$''$, which
corresponds to an expected time delay of a few weeks. The background
source is highly variable, and PKS1830--211 is thus a good candidate for
a measurement of the differential time delay between the cores.
Due to heavy extinction, the system can only be detected at wavelengths
longer than a few microns
At low radio frequencies, however, the observed continuum gets a
considerable contribution from the jet features. This contribution
must be subtracted in order to get the flux associated with the (variable)
cores. This requires the use of interferometers as well as modelling of
the emissivity distribution.
The jet--like features have a steep radio spectrum, improving
the situation at short wavelengths. At millimeter wavelengths
the flux from the jets are completely negligible and only the cores
contribute to the flux. The problem at millimeter wavelengths
is that angular separation of the cores can only be achieved using
interferometric techniques, which is both time consuming and difficult
given the northern location of millimeter interferometers.

Due to a fortunate configuration of obscuring molecular gas in the
lensing galaxy, the flux contributions from the NE and SW cores
can, however, easily be estimated using molecular absorption lines
and a single dish telescope with low angular resolution.
The lensing galaxy was actually detected through the presence of
millimetric molecular absorption lines, at a redshift
z$=$0.886 (Wiklind \& Combes 1996). The redshift of the background
source has recently been derived using IR spectroscopy and found to be
z$=$2.507 (Lidman et al. 1999). Several molecular absorption lines were
found to be saturated, among them the HCO$^+$(2--1) line, yet did not
reach zero intensity level.
Subsequent interferometer observations showed that the absorption
occurs only in front of the SW core component (Wiklind \& Combes 1998).
%A secondary weaker molecular absorption has now been found towards the
%NE component as well, separated in velocity by $-147$\,km\,s$^{-1}$
%(Wiklind \& Combes 1998).
The high opacity of the HCO$^+$ absortion line has been inferred through
the detection of isotopic variants, such as H$^{13}$CO$^+$ and H$^{18}$CO$^+$
(Wiklind \& Combes 1996, 1998). The depth of the absorption thus measures
directly the flux from the SW component, while the total continuum is the
sum of the fluxes from the SW and NE components (see Fig.\,1a).

\medskip

We have used the SEST 15m telescope to monitor the total continuum flux
and the depth of the HCO$^+$(2--1) line towards PKS1830--211 since April
1996 (Fig.\,1a). During this period the source has changed its flux by a
factor $\sim$2.5. Using the dispersion technique introduced by Pelt et al.
(1994) we derive a differential time delay of $24^{+5}_{-4}$ days, with
the NE component leading (Fig.\,1b. The result of Lovell et al. (1998),
based on low frequency interferometric techniques and model subtraction
of the jet components, are consistent with this value.
These two methods of deriving the time delay use different techniques and
their agreement gives additional confidence to the results.
In the lens model of Nair et al. (1993), our time delay measurement corresponds
to a Hubble constant $H_0 = 69^{+12}_{-11}$\,km\,s$^{-1}$\,Mpc$^{-1}$ (q$_0=0.5$).

\begin{figure}
\plotone{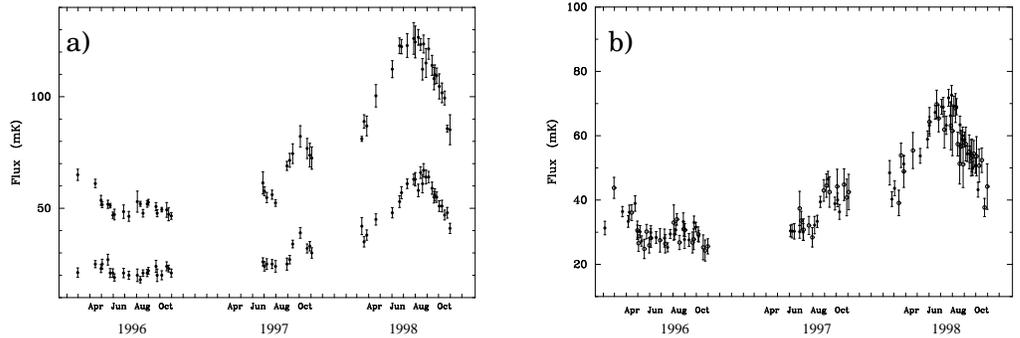}
\caption{{\bf a)} The total continuum flux (top) and the depth of
the HCO$^+$(2--1) absorption line (bottom).
{\bf b)}  The combined NE and SW flux components, where the NE flux has
been shifted by $-24$ days and corrected for a linear trend of the magnification}
\end{figure}

\acknowledgments

We are grateful to L.--{\AA}. Nyman, L. Haikala, F. Mac--Auliffe, F. Azagra,
A. Tieftrunk, F. Rantakyr\"{o} and K. Brooks for expert help with performing
the observations.

\end{document}